\documentclass[12pt]{iopart}
\usepackage[utf8]{inputenc}
\usepackage[T1]{fontenc}
\usepackage{amssymb}
\usepackage{graphicx}
\usepackage{xcolor}
\begin{document}

\title[Dual-frequency Doppler-free spectroscopy for simultaneous laser stabilisation]{Dual-frequency Doppler-free spectroscopy for simultaneous laser stabilisation in compact atomic physics experiments}

\author{Nathan Cooper$^1$, Somya Madkhaly$^1$, David Johnson$^1$, Daniele Baldolini$^1$ and Lucia Hackerm\"{u}ller$^1$}
\address{$^1$School of Physics and Astronomy, University of Nottingham, University Park, Nottingham, NG7 2RD, UK}

\ead{nathan.cooper@nottingham.ac.uk}

\vspace{10pt}
\begin{indented}
\item[]August 2021
\end{indented}

\begin{abstract}
Vapour cell spectroscopy is an essential technique in many fields; in particular, nearly all atom and ion trapping experiments rely on simultaneous spectroscopy of two atomic transitions, traditionally employing separate apparatus for each transition. 
Here, we demonstrate simultaneous spectroscopy on two atomic transitions using spatially-overlapped beams from two independent lasers, within a single spectroscopic apparatus. We show that, in addition to aiding compactness, this approach offers superior performance, leading to sharper spectroscopy peaks and stronger absorption signals. Doppler-free locking features become visible over a frequency range several hundred MHz wider than for standard saturated absorption spectroscopy. By exploring the full, 2D parameter space associated with dual-frequency spectroscopy, we reveal a lattice-like structure of sharp resonance features in 2D frequency space, which enhances experimental versatility by allowing laser frequency stabilisation anywhere within a wide manifold of locations in 2D frequency space. The process of simultaneous frequency stabilisation of two lasers is analysed in detail, revealing that it can be expected to produce significant improvements in laser stability when compared to conventional, single-frequency spectroscopy.
\end{abstract}

\section{Introduction}

Experiments employing cold, trapped atoms or ions underpin a wide range of fundamental research \cite{BECrev,spaceint,equivtest} and technological applications \cite{bize2005cold,gravimeter1,gravimeter2,MIT,sensor_review}. Such experiments require lasers that are accurately frequency-stabilised relative to specific atomic transitions, a function typically performed via feedback based on spectroscopy of a thermal atomic vapour. In addition, many important direct applications rely on atomic vapour spectroscopy --- for example the use of thermal atomic magnetometers in medical imaging \cite{SPMIC,medim2}, or of atomic spectroscopy to search for dark matter \cite{budker}. Improving the signal strength and the frequency sensitivity of atomic vapour spectroscopy is therefore beneficial across a wide range of applications and research areas.  

Frequency-stabilisation of lasers for atom cooling and trapping experiments is achieved via a feedback servo that controls the laser diode current and/or other feedback parameters. Typically at least two independent lasers are required, as light resonant with two different atomic transitions --- typically referred to as the ``cooler'' and ``repumper'' transitions --- is essential \cite{raab1987trapping}. The signal used for this feedback is  generated from spectroscopic measurements on an atomic vapour cell. A common technique for this is based on saturated absorption spectroscopy \cite{preston1996doppler,Demtroeder}, combined with modulation of the laser current and phase-sensitive detection of the spectroscopic signal \cite{riehle2006frequency}. 

Dual-frequency Doppler-free spectroscopy, in which two optical frequencies are overlapped on a single spatial path within an atomic vapour, can increase the strength and sharpness of Doppler-free resonance features \cite{dfds1,dfds2,dfds3,dfds4} and enhance frequency stabilisation of a single, modulated laser. 
We demonstrate that dual-frequency spectroscopy with two independent laser sources enables precise, simultaneous spectroscopy on the two transition manifolds relevant to atomic cooling and trapping experiments. The variation of both lasers' frequencies opens up a two-dimensional parameter space where optical pumping effects create a lattice of Doppler-free resonance features, offering an expanded set of potential ``locking points'' for laser frequency stabilisation. We explain how two lasers can be stabilised simultaneously based on the resulting spectroscopic signals, finding that the method enables both miniaturisation of the experimental apparatus and increased laser frequency stability. In \cite{optamot}, the technique is employed to stabilise two lasers simultaneously for the creation of a magneto-optical trap; here we provide a detailed analysis of the technique and its potential benefits for laser frequency stability. 

These results are of particular relevance to the burgeoning field of portable quantum technologies \cite{Blackett2, bongs2016uk}. Here, the achievable reductions in size, weight and complexity offered by allowing multiple beams to share one spatial pathway are important \cite{optamot}. Furthermore, the drive for miniaturisation will increase the desirability of using small vapour cells with correspondingly reduced optical depths \cite{minicell1,minicell2}, thus making enhancement of the strength and sensitivity of the locking signal yet more important. The ambition to operate these systems outside the laboratory, with potential exposure to increased environmental noise, also adds to the desirability of increasing signal strength.  

The paper is structured as follows: we give a brief overview of the experimental arrangement employed in this technique and present a rate-equation model that explains many of the observed key features of the approach. We then present experimental results spanning the full 2D frequency space associated with dual-frequency spectroscopy, before analysing the relative merits of this technique for laser stabilisation when compared to the status quo. 

\section{Setup and layout}

\begin{figure}[ht]
\centering
\includegraphics[width=\textwidth]{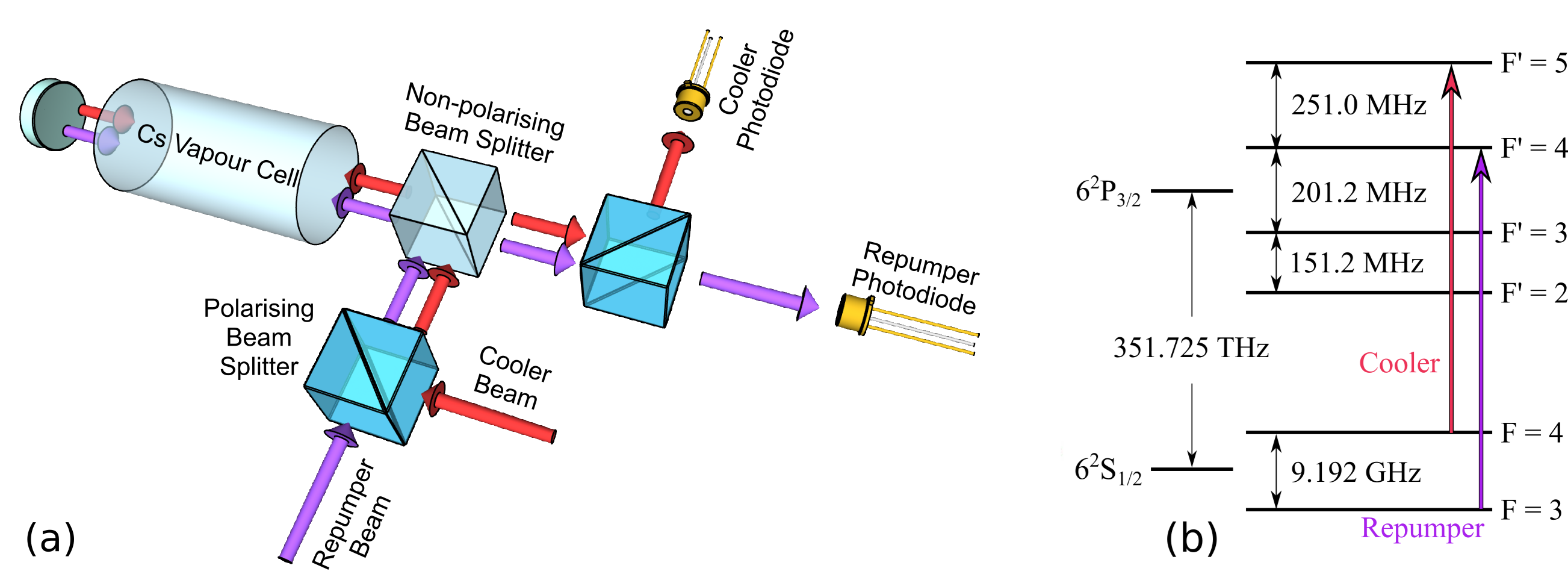}
\caption{(a): Experimental setup used for dual-beam spectroscopy. For clarity, the spatially-overlapped laser beams are illustrated as side-by-side in the figure. (b): Level structure for $^{133}$Cs, indicating the role of the `cooler' and `repumper' lasers.}
\label{fig:setup}
\end{figure}

We consider two orthogonally polarised laser beams from independent lasers, which are combined at a polarising beamsplitter. The two beams co-propagate through a Doppler-free spectroscopy setup \cite{Demtroeder}, sharing the same optical components, as illustrated in figure \ref{fig:setup}(a). This configuration is chosen 
as required for most atomic physics experiments, which rely on two frequencies (`cooler' and `repumper') for atom cooling, as shown for the example of $^{133}$Cs in Fig. \ref{fig:setup}(b).

The use of only a single optical frequency for spectroscopy on the D lines of alkali metal atoms results in pumping of the atomic population into a state not addressed by the pumping light, just as observed in magneto-optical trapping experiments \cite{raab1987trapping}, leading to a substantial attenuation of the atomic response and a weaker spectroscopic signal.
When light from two lasers is overlapped on the same spatial pathway in an atomic vapour, the absorption of each laser beam is affected by the presence of the other, leading e.g. to coherent effects such as electromagnetically-induced transparency \cite{eit} as well as bringing back parts of the atomic population into the cycling transition through `repumping'.
Simultaneous use of light resonant with transitions from both hyperfine states of the lower manifold (see figure \ref{fig:setup}(b)) can therefore be expected to improve spectroscopic signal strength.


Both beams have frequencies tuned close to resonance with the caesium D2 line: the `repumper' laser is resonant with transitions from the $F=3$ hyperfine state of the lower manifold to the upper manifold, and the `cooler' laser with transitions from the $F=4$ state of the lower manifold, matching the standard naming convention in the field --- see figure \ref{fig:setup}(b). After interacting with the cell the laser beams are separated at a polarising beamsplitter and their intensity is individually recorded on separate photodiodes shown in figure \ref{fig:setup}(a).

\begin{figure}[h!]
\centering
\includegraphics[width=\linewidth]{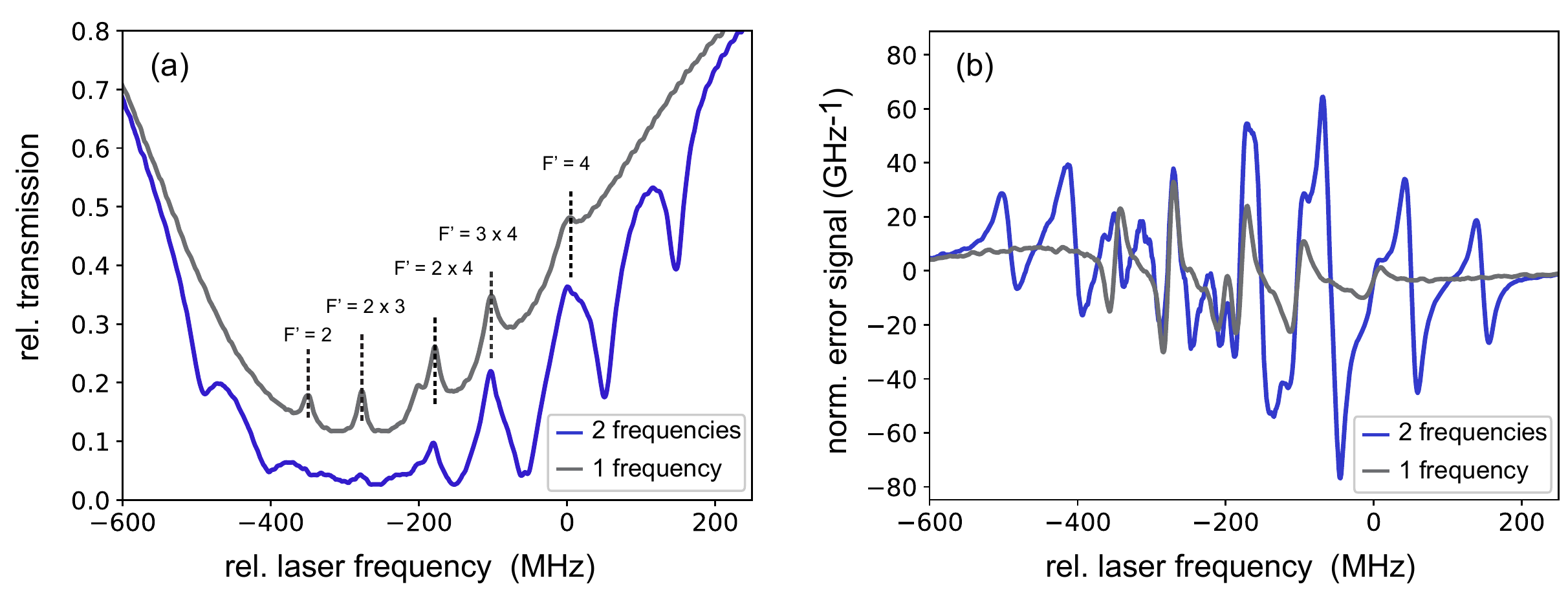}
\caption{(a): Spectroscopy signal from the repumper photodiode (see figure \ref{fig:setup}(a)) while light from the cooler laser, tuned $365$\,MHz below the $F=4 \rightarrow F'=5$ transition, is also present in the cell (purple line). A standard Doppler-free saturated absorption spectroscopy signal (grey line) is shown for reference. The addition of light tuned to the cooler transition substantially increases the size and spread of the Doppler-free features. (b): Normalised error signal (see text) resulting from the spectroscopic signals in (a). In both panels, the laser frequency axis is that of the repumper laser relative to the F=3$\rightarrow$F$'$=4 transition.}
\label{fig:cutthroughs}
\end{figure}

As an example, figure \ref{fig:cutthroughs} shows the results of a 1D experiment in which 0.37\,mW of cooler light, at a fixed frequency of -365\,MHz (relative to the F=4$\rightarrow$F$'$=5 transition), was directed into the apparatus alongside 0.25\,mW of repumper light, while the frequency of the repumper laser was scanned. The resulting absorption of the repumper light was determined by recording the output of the repumper photodiode from figure \ref{fig:setup}(a) and is shown in figure \ref{fig:cutthroughs}(a). For comparison, the figure also shows the result when the light from the cooler laser was blocked, which then corresponds to a standard saturated absorption spectroscopy signal. Fig. \ref{fig:cutthroughs}(a) clearly shows that the use of dual-frequency spectroscopy (blue line) substantially increases both the overall absorption of the light by the atomic vapour and the size and spread of Doppler-free resonance features in comparison to standard Doppler-free spectroscopy with only a single laser beam (grey line). Figure \ref{fig:cutthroughs}(b) shows the corresponding ``normalised error signal,'' which is defined as the fractional change in light transmission per unit frequency, given by $T^{-1}\left(dT/dF_r \right)$, where $T$ is the fraction of the repumper light that is transmitted through the atomic vapour and $F_r$ is the frequency of the repumper laser. This derivative represents the `ideal' form of the error signal typically employed for spectroscopic feedback stabilisation of a laser's frequency. The comparison in Figure \ref{fig:cutthroughs}(b) displays a steeper gradient around the locking points when two frequencies are present as well as an increased number of potential locking points; and related to that a wider frequency space over which lock points are available. 
The ``normalised error signal'' demonstrates the relevance of this method with respect to laser frequency stabilisation. The increase in the maximum gradient of the error signal will improve frequency stability when applied for feedback stabilisation of a laser, and the increase in the amplitude of the features will improve the robustness of laser stabilisation in noisy environments by offering an increased ``capture range'' about the desired stabilisation frequency.


\section{Model and Theory}

An intuitive analysis of this system can explain the enhanced absorption and the increased number of absorption peaks. We will also see, in panel (a) of figures \ref{fig:signals_r} and \ref{fig:signals_c}, that a lattice-like structure is observed in 2D frequency space; the same intuitive analysis of optical pumping effects within a thermal velocity distribution can explain this phenomenon.  
Consider, for example, the absorption of the cooler light. When the repumper light is on resonance with transitions in the same atoms, within the thermal atomic velocity distribution, as the light from the cooler beam, it enhances the absorption of the cooler light. For co-propagating beams the simultaneous resonance condition requires that the two laser frequencies are separated by a fixed amount, thus creating a set of sharp features, similar to those seen in saturated absorption spectroscopy, that map out diagonal lines of positive slope in 2D frequency space. For counter-propagating beams, the same effect is present but the sign of the slope is reversed. Since both co-propagating and counter-propagating beams are present in the vapour cell, the dual-frequency optical pumping effects can be expected to produce diagonal line features with both positive and negative slope. 

To go beyond this intuitive description, we develop a simple theoretical model that correctly predicts the key features of our results and elucidates the underlying physical mechanisms. 

We approximate the transient behaviour of atoms as they traverse the beam in the following way: we assume that some fraction of the atoms, $F_N$, have recently entered the laser beam and remain in an equal mixture of the two hyperfine states of the 6S$_{1/2}$ level. The remaining atomic population is assumed to be in a state of dynamic equilibrium. This is likely to be a reasonable approximation, as very few cycles of optical pumping are required to redistribute the atomic population.  

In order to determine the influence of these `equilibrium state' atoms on the spectroscopic signals, we adopt a similar approach to that presented in previous work on atom trapping using optical pumping effects \cite{moptrap}, developing a rate-equation based model in which we consider the six-level system shown in figure \ref{fig:setup}(b). For convenience, we label the atomic states A-F, as shown in the figure. For now we consider only a single atom with fixed laser detunings --- the thermal distribution of atomic velocities and corresponding Doppler shifts will be factored in later. We define a set of rate coefficients, $\tau_{\mathrm{ij}}$ and $\Gamma_{\mathrm{ij}}$, such that the stimulated and spontaneous transition rates between, for example, states E and B are given by $\tau_{\mathrm{EB}}I_{\mathrm{EB}}$ and $\Gamma_{\mathrm{EB}}$ respectively, where $I_{\mathrm{EB}}$ is the intensity of the laser light tuned to the relevant set of transitions (i.e. cooler or repumper). The spontaneous decay rates for the relevant transitions are already known accurately --- see for example \cite{steck}. To determine the rate coefficients for stimulated transitions, we equate the steady-state results for the upper state population produced by our rate equation model to those produced by solving the full optical Bloch equations for a two level system. For a transition with spontaneous decay rate $\Gamma$, illumination of detuning $\delta$ and intensity $I$, with Rabi frequency $\Omega$, we obtain


\begin{equation}
\frac{\Omega^{2}/4}{\delta^{2}+\Omega^{2}/2+\Gamma^{2}/4} = \frac{\tau I}{2\tau I+\Gamma}.
\end{equation}

Therefore, labeling the dipole matrix element $\langle \mathrm{E}|x|\mathrm{B} \rangle$ between two levels as $X_{\mathrm{EB}}$, we find that

\begin{equation}
\tau_{\mathrm{EB}} = \frac{\Omega_{\mathrm{EB}}^{2} \Gamma_{\mathrm{EB}}}{4 I_{\mathrm{EB}}(\delta_{\mathrm{EB}}^{2}+\Gamma_{\mathrm{EB}}^{2}/2)} = \frac{e^{2} \left| X_{\mathrm{EB}} \right|^{2} \Gamma_{\mathrm{EB}}}{2 \hslash^{2} c \epsilon_{0} (\delta_{\mathrm{EB}}^{2}+\Gamma_{\mathrm{EB}}^{2}/2)}
\label{transcoeffs}
\end{equation}

and similar results for the other transitions. Hence, using each state's label to denote the population fraction in that state, the rate equations governing the system are given by:

\begin{equation}
\label{rate1}
\frac{\mathrm{dF}}{\mathrm{dt}} = (\mathrm{B}-\mathrm{F})\tau_{\mathrm{FB}}I_{\mathrm{FB}}-\mathrm{F}\Gamma_{\mathrm{FB}},
\end{equation}

and corresponding expressions for the time derivatives of the populations of the remaining states (see Supplementary Materials).
Setting all time derivatives to zero and the total fractional population across all states equal to 1, we solve the equations to find the fractional steady-state populations in each state. The full derivation is given in the supplementary material. Here we simply quote the result, first defining the following notation:

\begin{equation}
    \gamma_{\mathrm{ij}} = \tau_{\mathrm{ij}}I_{\mathrm{ij}}+\Gamma_{\mathrm{ij}},
\end{equation}

\begin{equation}
    k_{\mathrm{ij}} = \frac{\tau_{\mathrm{ij}}I_{\mathrm{ij}}}{\sum_{\mathrm{k}} \gamma_{\mathrm{ik}}},
\end{equation}






\begin{equation}
    \zeta_{\mathrm{i}} = \frac{1}{\sum_{\mathrm{j}} \tau_{\mathrm{ji}}I_{\mathrm{ji}}},
\end{equation}

and the composite coefficient

\begin{equation}
    C_1 = \frac{\left( k_{\mathrm{EA}}\gamma_{\mathrm{EB}}+k_{\mathrm{DA}}\gamma_{\mathrm{DB}}\right) \zeta_{\mathrm{B}}}{1-\left( k_{\mathrm{FB}}\gamma_{\mathrm{FB}}+k_{\mathrm{EB}}\gamma_{\mathrm{EB}}+k_{\mathrm{DB}}\gamma_{\mathrm{DB}}\right) \zeta_{\mathrm{B}}},
\end{equation}

where the summations are carried out over all dipole-allowed transitions. Note that repeated indices do not imply summation here --- summations are used only where explicitly stated. We now find that in the steady state

\begin{equation}
    A = \left[ 1+\sum_{\mathrm{i}} k_{\mathrm{iA}} + C_1 \left(1+ \sum_{\mathrm{j}} k_{\mathrm{jB}}\right) \right] ^{-1},
\end{equation}

with the remaining fractional populations given by

\begin{equation}
    B = C_1 A,
\end{equation}

and for the upper manifold

\begin{equation}
    i = k_{\mathrm{iA}} A + k_{\mathrm{iB}} B.
\end{equation}

Having obtained the steady state populations we can now determine the rate of photon loss per equilibrium-state atom, $L_{eq}$, from a given beam as

\begin{equation}
    L_{eq} = \sum (j-i)\tau_{\mathrm{ij}}I_{\mathrm{ij}},
    \label{lossrate}
\end{equation}

where the sum is taken over all combinations of upper manifold states $i$ and lower manifold states $j$ between which dipole-allowed transitions can be stimulated by the chosen beam (ignoring the negligible stimulation of cooler/repumper transitions by repumper/cooler lasers respectively). The un-pumped atoms can be taken into consideration at this stage as well, yielding

\begin{equation}
    L = (1-F_N) \sum (j-i)\tau_{\mathrm{ij}}I_{\mathrm{ij}}+F_N \sum \frac{1}{2}\tau_{\mathrm{ij}}I_{\mathrm{ij}}.
    \label{lossrate}
\end{equation}

Equation (\ref{lossrate}) can be used to determine the rate of attenuation of a laser beam by stationary atoms subject to known illumination conditions. In order to accurately model a thermal atomic vapour, the velocity distribution of the atoms and corresponding Doppler shift in each beam's effective detuning must be taken into account. This can be done by integrating equation (\ref{lossrate}) over the atomic velocity distribution, where the dependence of $L$ on atomic velocity comes in via the dependence of the values of $\delta_{\mathrm{ij}}$ in equation (\ref{transcoeffs}) on atomic velocity (due to the Doppler shift) and the corresponding variation in the values of $\tau_{\mathrm{ij}}$. We define a new variable, $L_{\mathrm{thermal}}$, as the average loss rate of photons from the beam per atom, given the atoms' thermal velocity distribution. In the case of a thermal atomic vapour at temperature $T$, considering only the first-order Doppler shift, this is given by

\begin{equation}
    L_{\mathrm{thermal}} = \frac{1}{N} \int _{-\infty}^{\infty} L(v) \exp{\left(-mv^2/2k_{\mathrm{B}}T\right)} dv, \label{ltherm}
\end{equation}

where $m$ is the mass of the atoms, $k_{\mathrm{B}}$ is the Boltzmann constant and the integration variable $v$ corresponds to the atomic velocity component along the axis of the laser beams. $N$ is the normalisation constant for the 1D Boltzmann distribution. Furthermore, we must consider that there are counter-propagating beams within the vapour cell. These can be taken into account simply by summing the contributions of the different beams to the stimulated transition rates, such that in the equations above $\tau_{\mathrm{ij}}I_{\mathrm{ij}}$ becomes $\tau_{\mathrm{ij}}I_{\mathrm{ij}}(\mathrm{beam 1})+\tau_{\mathrm{ij}}I_{\mathrm{ij}}(\mathrm{beam 2})$. Note that for counter-propagating beams the sign of the Doppler shift on the value of $\delta_{\mathrm{ij}}$ in equation (\ref{transcoeffs}) will be opposite for the two beams. With this substitution made equation (\ref{ltherm}) can then be applied individually to each laser beam present.

Equation (\ref{ltherm}) can now be used to deduce the mean optical absorption cross-section per atom as a function of both laser frequencies. 
The results of this are shown alongside our experimental data in figures \ref{fig:signals_r} and \ref{fig:signals_c}. The expected key features can clearly be seen: the presence of each beam increases the strength of the absorption of the other beam and creates sharp locking features, similar to those seen in saturated absorption spectroscopy, in a diagonal `criss-cross' pattern across the whole Doppler valley. While the output of the model does not exactly match the experimental data, this is expected given the assumptions made in the model --- a much more complex model would be necessary to show quantitative agreement with experimental results, but ours provides a useful guide to the overall trends in the system and the nature of the observed features.

\section{Experimental results}

\begin{figure}[h!]
\centering
\includegraphics[width=\linewidth]{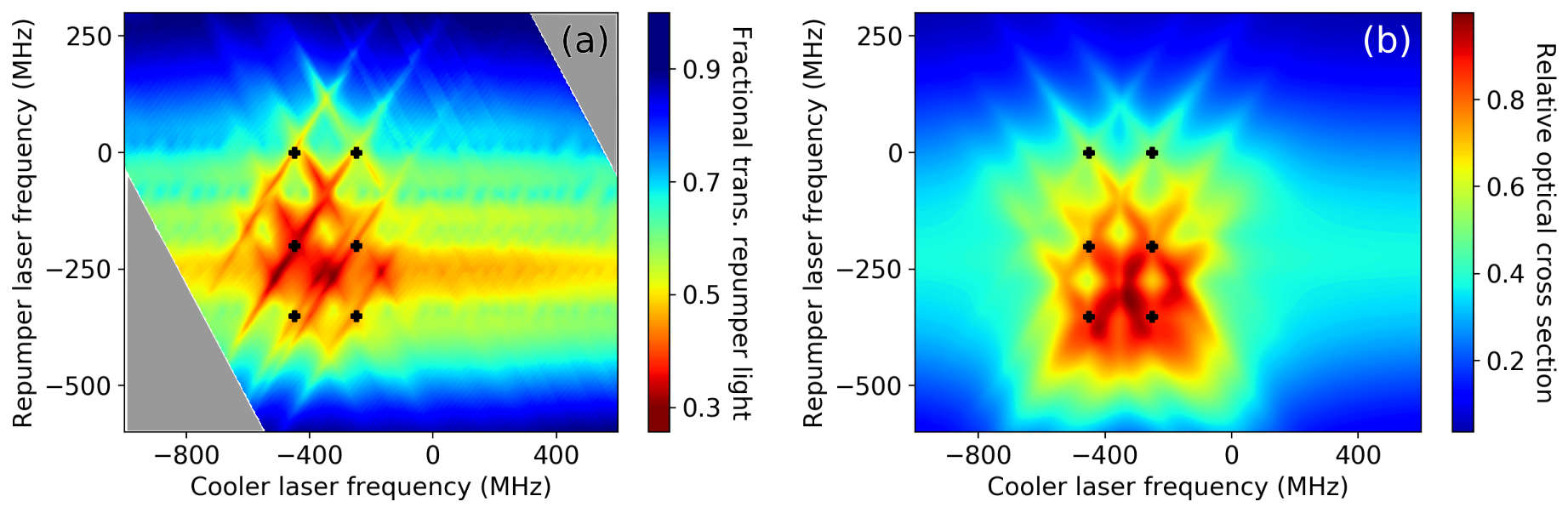}
\caption{Spectroscopic signal from the repumper laser photodiode (a) as a function of the frequencies of both lasers. See text for details. Frequencies for the cooler and repumper lasers are given relative to the F=4$\rightarrow$F$'$=5 and F=3$\rightarrow$F$'$=4 transitions respectively. Panel (b), shows the prediction of our rate-equation model for the relative optical absorption cross-section per atom from the repumper beam. To guide the eye, black `+' symbols indicate the points where the two lasers are simultaneously resonant with relevant transitions in stationary atoms, such that optical pumping effects lead to diagonal line features. The the rate equation model describes the observed key features very well. The simulation assumes that the intensity of the return beam is always half of that of the in going beam.}
\label{fig:signals_r}
\end{figure}

\begin{figure}[h!]
\centering
\includegraphics[width=\linewidth]{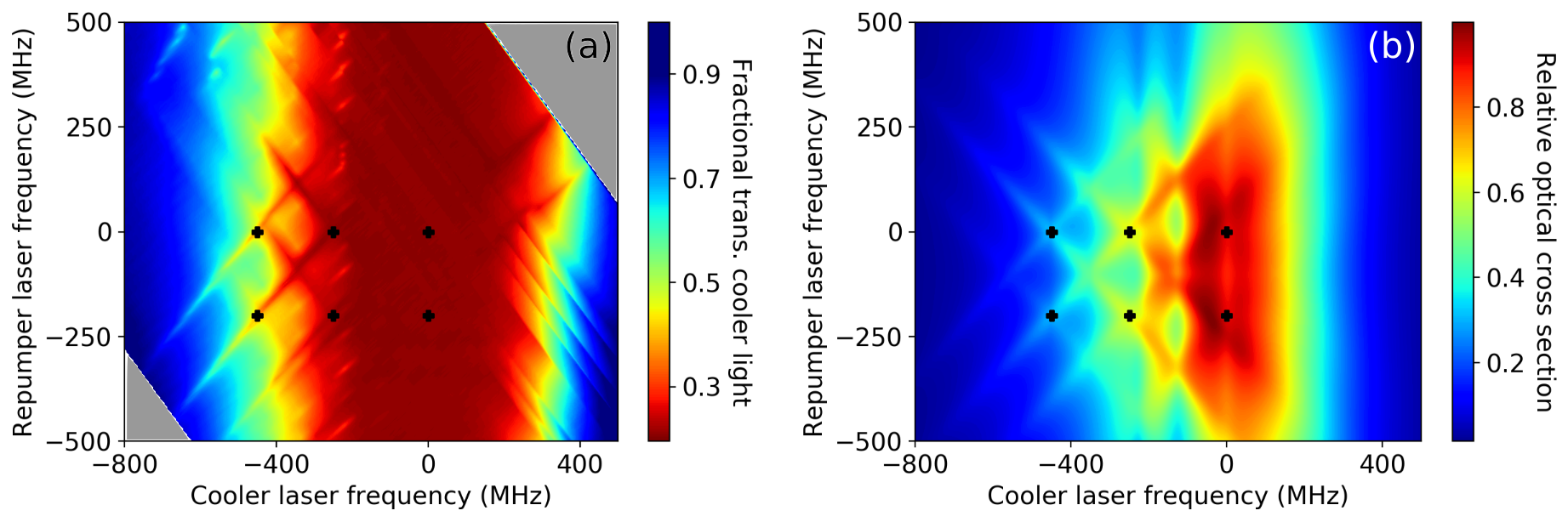}
\caption{Spectroscopic signal from the cooler laser photodiode (a) as a function of the frequencies of both lasers. See figure \ref{fig:setup} and main text for details. 
Shown alongside, in panel (b), is the prediction of our rate-equation model (see above) for the relative optical absorption cross-section per atom from the cooler beam.} 
\label{fig:signals_c}
\end{figure}

Figures \ref{fig:signals_r} and \ref{fig:signals_c} show the output signals from each of the photodiodes shown in figure \ref{fig:setup}(a), as a function of the frequencies of both the cooler and repumper lasers. These were obtained by synchronously scanning both laser frequencies across the relevant frequency range. This was done by adding a linear ramp to the current supplied to each laser diode, and by simultaneously ramping the voltage supplied to piezoelectric transducers that control the alignment of diffraction gratings used for external cavity feedback. The result is laser frequencies described by the equations
\begin{equation}
    F_1 = A+Bt ~~~~~ \mathrm{and} ~~~~~ F_2 = C+Dt,
\end{equation}
where $t$ is the time since the start of the ramp, $F_1$ and $F_2$ are the frequencies of the cooler and repumper lasers, respectively, and $A$-$D$ are constant coefficients. The resulting equation expressing $F_2$ as a function of $F_1$ is
\begin{equation}
    F_2 = \frac{D}{B}F_1 + \left(C-\frac{DA}{B} \right).
\end{equation}
This equation clearly describes a diagonal line in the parameter-plane displayed in figure \ref{fig:signals_r}, with gradient $D/B$ and offset C-(DA/B). By adjusting either of the static frequency offsets, $A$ and $C$, it is then possible to collect data along multiple such lines and build up a full, 2D dataset as displayed. Note that the boundaries of the region within which data was collected consequently form diagonal lines in 2D frequency space --- hence the greyed-out triangles in the corners of figures \ref{fig:signals_r} and \ref{fig:signals_c}. An absolute frequency reference was provided by simultaneously directing light from each of the lasers to a standard saturated absorption spectroscopy apparatus; this enables independent confirmation of each laser's frequency via a well established technique. The diameter of each beam was 1.25\,mm and the powers used were 0.14\,mW for the cooler laser and 0.23\,mW for the repumper laser.  

The results in figures \ref{fig:cutthroughs}, \ref{fig:signals_r} and \ref{fig:signals_c} clearly show that the presence of the additional cooler/repumper frequency light enhances the absorption signal of the repumper/cooler light by the atomic vapour; figure \ref{fig:cutthroughs} represents a vertical slice through figure \ref{fig:signals_r}(a), while conventional spectroscopy corresponds to an equivalent slice in the limit of a far-detuned cooler laser, which is approached at the boundaries of the cooler-frequency axis in figure \ref{fig:signals_r}(a). An enhancement of the absorption signal translates to more accurate laser frequency stabilisation by providing a feedback signal with improved signal to noise ratio. 
In addition, figures \ref{fig:signals_r} and \ref{fig:signals_c} also demonstrate that sharp response features are created in a diagonal grid pattern over a frequency range of about 700 MHz --- considerably broader than the 450 MHz (for the cooler laser) or 350 MHz (for the repumper laser) over which conventional saturated absorption spectroscopy produces Doppler-free resonance features. This enables the technique to be used for frequency stabilisation at a much wider range of frequency offsets than is usually possible.

\section{Application to laser frequency stabilisation}

We focus on simultaneous frequency stabilisation of two lasers using the same apparatus. The technique could also be of use for enhanced frequency stabilisation of a single laser via the injection of light from an independently stabilised laser --- this possibility is considered in detail in the Supplementary Material.

An appropriate figure of merit is the factor by which the frequency deviations of a laser stabilised via dual-frequency spectroscopy could be reduced, compared to those of a laser stabilised using conventional techniques, under the assumption that the feedback hardware works perfectly and stability is limited entirely by the quality of the spectroscopic signal. We shall refer to this as the ``signal-limited stability factor'' (SLSF). This is an appropriate figure of merit for characterisation of the spectroscopic technique being used; other measures, such as a direct measurement of a stabilised laser's frequency stability, can conflate the efficacy of the spectroscopic technique with other issues, such as the finite bandwidth of feedback hardware or electronic noise in the feedback circuitry.


To determine numerical values for the SLSF, we must consider a specific laser stabilisation scheme. We examine the case of saturated absorption spectroscopy, combined with laser current modulation and phase-sensitive detection of the corresponding spectroscopic signal \cite{riehle2006frequency}. Note that dual-frequency spectroscopy is not limited to use with this method and similar results are expected for alternative schemes, such as dichroic atomic vapour laser locking \cite{davll,millett2006davll} or polarisation spectroscopy \cite{polspec,harris2006polarization}. 

Locking based on saturated absorption spectroscopy and current modulation generates a response signal, $S$, approximately proportional to the rate of change of the saturated absorption signal voltage with laser frequency:
\begin{equation}
    S \approx C \frac{dV}{dF},
\end{equation}
where $C$ is a system-specific constant, $F$ is the laser frequency and $V$ is the normalised output voltage of the photodiode. 
The parameter of interest --- the sensitivity of the response signal to changes in laser frequency --- is therefore given by
\begin{equation}
    \frac{dS}{dF} \approx C \frac{d^2V}{dF^2}.
\end{equation}

For simultaneous stabilisation of two lasers, composite feedback signals must be generated that depend (at least to first order about the desired `lock point' - the position in 2D frequency space where it is intended to stabilise both lasers) only on the frequency of one laser. We consider two lasers with frequencies $F_1$ and $F_2$ generating corresponding photodiode outputs $V_1$ and $V_2$. We assume that the currents of the two lasers are modulated at different frequencies to avoid direct cross-talk. 

Assuming that all derivatives are evaluated at the chosen locking point, the gradient of the spectroscopic signal after demodulation, $S_1$, is given by 
\begin{equation}
    \frac{dS_1}{dF_1} = C_1\frac{d^2V_1}{dF_1^2}
\end{equation}
and
\begin{equation}
    \frac{dS_1}{dF_2} = C_1\frac{d^2V_1}{dF_1~dF_2},
\end{equation}
with corresponding expressions for the gradient of $S_2$. We can therefore define a composite parameter $\chi_1$ with no first-order dependence on $F_2$:
\begin{equation}
    \chi_1 = S_1-\upsilon_1 S_2,
\end{equation}
where
%
%
\begin{equation}
    \upsilon_1 = \frac{C_1}{C_2}  \frac{d^2V_1}{dF_1~dF_2}  \left(  \frac{d^2V_2}{dF_2^2}  \right)^{-1}.
    \label{nu}
\end{equation}
Corresponding expressions obviously exist for $\chi_2$, following the same derivation. Thus it is possible to generate a feedback signal for each laser that is independent of the frequency of the other laser. This makes it possible to stabilise both lasers simultaneously based on the generated feedback signals. 

\begin{figure}[h!]
\centering
\includegraphics[width=0.7\linewidth]{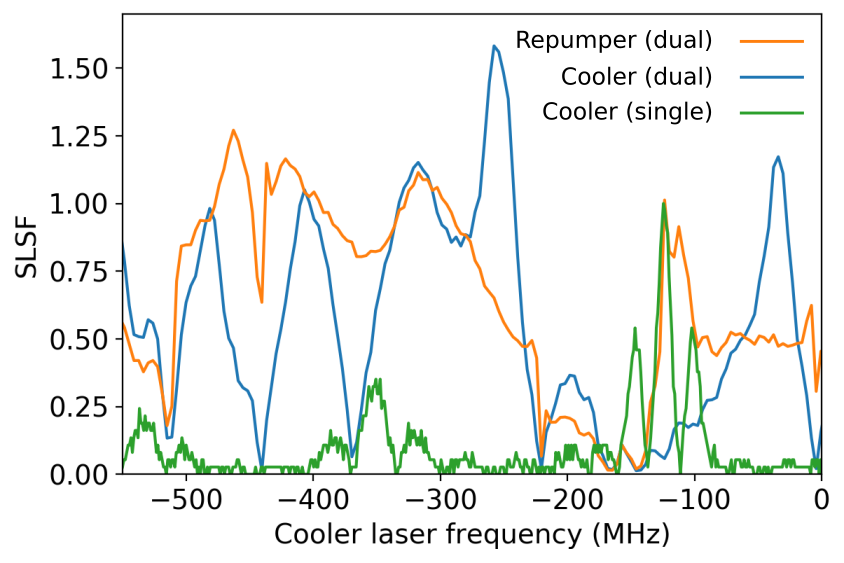}
\caption{Signal-limited stability factor (SLSF --- see text for full definition) resulting from laser stabilisation based on dual-frequency spectroscopy, compared to the maximum achievable using conventional saturated absorption spectroscopy with only a single frequency component. The repumper laser frequency is set to -370 MHz and the SLSF is determined by applying equation (\ref{duores}) to our data. It can be seen that dual-frequency spectroscopy offers generally improved stability, over a wider range of frequencies, than conventional spectroscopy. In addition, multiple locations exist at which the stability of \emph{both} lasers is simultaneously greater than the maximum achievable with single-frequency spectroscopy. Note that this figure displays only a 1D slice through 2D frequency space; the broader manifold of locations that can be accessed by tuning both laser frequencies offers high stability at an even greater range of lock points. Frequencies for the cooler and repumper lasers are given relative to the F=4 to F$'$=5 and F=3 to F$'$=4 transitions respectively.}
\label{fig:dualstab}
\end{figure}

We now determine the expected laser frequency stability and compare it to that achieved with conventional saturated absorption spectroscopy. The SLSF is equal to the ratio of the sensitivity of the spectroscopic signal (to changes in laser frequency, about the desired lock point) in dual-frequency spectroscopy to the same parameter in a conventional spectroscopic setup; in this case the SLSF for laser 1 (`$E_1$') is therefore given by
\begin{equation}
    E_1 = \left. \frac{d\chi_1}{dF_1} \right/ \left( \frac{dS_{\mathrm{con}}}{dF_{\mathrm{con}}} \sqrt{1+\nu_1^2} \right),
\end{equation}
where $dS_{\mathrm{con}}/dF_{\mathrm{con}}$ is the gradient of the feedback signal about the lock point in an equivalent conventional saturated absorption spectroscopy apparatus. The factor of $\sqrt{1+\nu_1^2}$ normalises against the amplification of the feedback signal that has been performed via post-processing, and consequently therefore amplifies the experimental noise as well as the desired signal and cannot produce an improvement in frequency stability. We thus find that 
\begin{equation}
    \label{duores}
    E_1 = \left. \left( \frac{dS_1}{dF_1}-\upsilon_1 \frac{dS_2}{dF_1} \right) \right/ \left( \frac{dS_{\mathrm{con}}}{dF_{\mathrm{con}}} \sqrt{1+\nu_1^2} \right),
\end{equation}
and the equivalent expression for $E_2$. As an example, the SLSF was calculated for a 1D subset of the experimental data shown in figures \ref{fig:signals_r} and \ref{fig:signals_c}, for both cooler and repumper lasers. This is displayed in figure \ref{fig:dualstab}; for illustrative purposes, the SLSF of conventional spectroscopy (which is by definition equal to 1 at the optimum locking point for single-frequency spectroscopy) is also plotted. Here, the repumper laser frequency is set 370 MHz below the F=3 to F$'$=4 transition and the cooler laser frequency is varied. It can be seen that there are multiple locations in frequency space where the SLSF exceeds 1 for both lasers simultaneously. This shows that, even when stabilising two lasers using the same apparatus, and when considering only the laser for which the result is least favourable, dual-frequency spectroscopy still offers better performance than conventional Doppler-free spectroscopy.

It is worth noting that, by selecting an appropriate lock point in 2D frequency space, the technique allows one to prioritise the frequency stability of one laser, further enhancing performance for one laser at the expense of worse performance for the other. This may be useful in situations where specific experimental applications have asymmetric frequency tolerances for the two lasers --- consider, for example, the very different tolerances of a MOT to frequency deviations of the cooler and repumper lasers \cite{raab1987trapping}. 

\section{Conclusions}

Our results show that the use of spatially overlapping beams tuned to different atomic transitions can allow optical pumping effects to be exploited to enhance signal strength in atomic vapour spectroscopy. The technique also allows the two beams to share the same optical path, thus reducing total system size and component usage, as well as allowing laser stabilisation at a greater range of frequencies than is possible via conventional spectroscopic techniques. In our experimental arrangement, which generates a feedback signal for laser stabilisation to the D2 line of Caesium via laser current modulation and demodulation of the corresponding spectroscopic signal at the same frequency, we found that a single apparatus could perform spectroscopy on two lasers simultaneously, while also producing a feedback signal that was more sensitive to frequency deviations of either laser than that generated using only one laser. 

The technique is likely to be of use wherever space, weight and optical components are at a premium as well as in situations where signal strength and frequency sensitivity are important. This will range from portable devices where reductions in size and weight are important, through to precision lab-based experiments where the accuracy of laser frequency stabilisation is paramount. 

Future extensions of this technique could include demodulation of the signal from one photodiode at two different frequencies --- corresponding to the two laser current modulation frequencies. This would allow measurement of, for example, $dV_1/dF_1$ and $dV_1/dF_2$ from a single photodiode; in the situation considered above $dV_1/dF_1$ is obtained by demodulation of the signal from one photodiode and $dV_2/dF_2$ from the other. In principle this could allow for both lasers to be stabilised using only one photodiode, or alternatively the extra information obtained by using two photodiodes, each demodulated at two different frequencies and thus generating four distinct feedback signals, could allow for yet more accurate stabilisation of both lasers. Two-laser forms of modulation transfer spectroscopy \cite{modulation_transfer} should also be possible, allowing this important technique to be exploited through current modulation of either one of the two lasers, avoiding the need for costly electro-optic modulation equipment.

\section*{References}
\bibliographystyle{iopart-num}
\bibliography{sample}

\section*{Acknowledgements}
This work was supported by IUK project No.133086 and the EPSRC grants EP/R024111/1 and EP/M013294/1 and by the European Comission grant ErBeStA (no. 800942). The authors thank Bethany Foxon and Igor Lesanovsky for useful discussions.



\section*{Additional information}

The authors declare the following competing interests: N.C. and L.H. are inventors on UK pending patent application GB 1916446.6 (applicant: University of Nottingham, inventors: Nathan Cooper, Lucia Hackerm\"{u}ller, Laurence Coles) for a miniaturised spectroscopy device in which dual-frequency beam overlap is exploited as an aid to compactness. All data necessary to support the conclusions of this article are given in the article. All further data related to this work are available from the authors upon request.

\end{document}


\title[Supplementary material for: Dual-frequency Doppler-free spectroscopy]{Supplementary material for: Dual-frequency Doppler-free spectroscopy for simultaneous laser stabilisation in compact atomic physics experiments}

\author{Nathan Cooper$^1$, Somya Madkhaly$^1$, David Johnson$^1$, Daniele Baldolini$^1$ and Lucia Hackerm\"{u}ller$^1$}
\address{$^1$School of Physics and Astronomy, University of Nottingham, University Park, Nottingham, NG7 2RD, UK}

\ead{nathan.cooper@nottingham.ac.uk}

\vspace{10pt}
\begin{indented}
\item[]August 2021
\end{indented}

\begin{abstract}
Herein we give the full derivation of the solutions of our rate-equation model of the pumped atomic vapour system. We also provide full details of our method of numerical estimation of second derivatives --- the method is not novel or of inherent interest, but the exact details of the estimation method used can affect the final results displayed in figure 4 of the main article, and we therefore provide them for completeness. We briefly review some of the complications that would be involved in constructing a full theoretical model of the system, capable of accurately reproducing experimental results. Though such a model would be complex and unlikely to reveal new fundamental physics, it does have significant practical applications and is intended to be the subject of future research. We present the data from figure 3(a) of the main article on a logarithmic scale, which illustrates how the technique is likely to be even more effective in miniaturised devices that contain only small vapour cells, with a correspondingly lower optical depth. We discuss the likely efficacy of the technique for enhanced frequency-stabilisation of a single laser via the injection of light from an independently stabilised source. Finally, we consider how the technique is likely to be of particular use in experiments involving atomic Lithium.
\end{abstract}

\flushbottom
\maketitle
%
%
\thispagestyle{empty}

\section{Solution to rate equations}

First, we give the expressions for the time derivatives of the remaining atomic state populations:

\begin{equation}
\frac{\mathrm{dE}}{\mathrm{dt}} = (\mathrm{A}-\mathrm{E})\tau_{\mathrm{E}\mathrm{A}}I_{\mathrm{E}\mathrm{A}}+(\mathrm{B}-\mathrm{E})\tau_{\mathrm{E}\mathrm{B}}I_{\mathrm{E}\mathrm{B}}-\mathrm{E}\Gamma_{\mathrm{E}\mathrm{B}}-\mathrm{E}\Gamma_{\mathrm{E}\mathrm{A}},
\end{equation}

\begin{equation}
\frac{\mathrm{dD}}{\mathrm{dt}} = (\mathrm{B}-\mathrm{D})\tau_{\mathrm{D}\mathrm{B}}I_{\mathrm{D}\mathrm{B}}+(\mathrm{A}-\mathrm{D})\tau_{\mathrm{D}\mathrm{A}}I_{\mathrm{D}\mathrm{A}}-\mathrm{D}\Gamma_{\mathrm{D}\mathrm{B}}-\mathrm{D}\Gamma_{\mathrm{D}\mathrm{A}},
\end{equation}

\begin{equation}
\frac{\mathrm{dC}}{\mathrm{dt}} = (\mathrm{A}-\mathrm{C})\tau_{\mathrm{C}\mathrm{A}}I_{\mathrm{C}\mathrm{A}}-\mathrm{C}\Gamma_{\mathrm{C}\mathrm{A}},
\end{equation}

\begin{equation}
  \fl \frac{\mathrm{dB}}{\mathrm{dt}} = (\mathrm{F}-\mathrm{B})\tau_{\mathrm{FB}}I_{\mathrm{FB}}+(\mathrm{E}-\mathrm{B})\tau_{\mathrm{EB}}I_{\mathrm{EB}}+(\mathrm{D}-\mathrm{B})\tau_{\mathrm{D}\mathrm{B}}I_{\mathrm{D}\mathrm{B}}+\mathrm{F}\Gamma_{\mathrm{F}\mathrm{B}}+\mathrm{E}\Gamma_{\mathrm{EB}}+\mathrm{D}\Gamma_{\mathrm{D}\mathrm{B}},
\end{equation}

and

\begin{equation}
\label{rate2}
\fl \frac{\mathrm{dA}}{\mathrm{dt}} = (\mathrm{E}-\mathrm{A})\tau_{\mathrm{EA}}I_{\mathrm{EA}}+(\mathrm{D}-\mathrm{A})\tau_{\mathrm{DA}}I_{\mathrm{DA}}+(\mathrm{C}-\mathrm{A})\tau_{\mathrm{C}\mathrm{A}}I_{\mathrm{C}\mathrm{A}}+\mathrm{E}\Gamma_{\mathrm{EA}}+\mathrm{D}\Gamma_{\mathrm{D}\mathrm{A}}+\mathrm{C}\Gamma_{\mathrm{C}\mathrm{A}}.
\end{equation}

Then, employing the terms defined in equations (1)-(2) and (4)-(7) of the main article, and setting all time derivatives to zero, as is the case in the steady state, equation (3) of the main text, together with its counterparts above, can then be re-arranged to give

\begin{equation}
    F = \frac{B \tau_{FB} I_{FB}}{\gamma_{FB}} = B k_{FB},
    \label{F}
\end{equation}

\begin{equation}
    E = \frac{A\tau_{EA}I_{EA}+B\tau_{EB}I_{EB}}{\gamma_{EA}+\gamma_{EB}} = A k_{EA}+Bk_{EB},
\end{equation}

\begin{equation}
    D = \frac{A\tau_{DA}I_{DA}+B\tau_{DB}I_{DB}}{\gamma_{DA}+\gamma_{DB}}= A k_{DA}+Bk_{DB},
    \label{D}
\end{equation}
\begin{equation}
    C = \frac{A\tau_{CA}I_{CA}}{\gamma_{CA}}= B k_{CA},
    \label{C}
\end{equation}

\begin{equation}
    B = \frac{D\gamma_{DB}+E\gamma_{EB}+F\gamma_{FB}}{\tau_{DB}I_{DB}+\tau_{EB}I_{EB}+\tau_{FB}I_{FB}} = \frac{D\gamma_{DB}+E\gamma_{EB}+F\gamma_{FB}}{\zeta_B},
    \label{B}
\end{equation}

and

\begin{equation}
    A = \frac{C\gamma_{CA}+D\gamma_{DA}+E\gamma_{EA}}{\tau_{CA}I_{CA}+\tau_{DA}I_{DA}+\tau_{EA}I_{EA}} = \frac{C\gamma_{CA}+D\gamma_{DA}+E\gamma_{EA}}{\zeta_A}.
\end{equation}

Substitution of eqns (\ref{F}) to (\ref{D}) into (\ref{B}) then yields

\begin{equation}
    B = \frac{\gamma_{DB}(A k_{DA}+Bk_{DB})+\gamma_{EB}(A k_{EA}+Bk_{EB})+\gamma_{FB}Bk_{FB}}{\zeta_B}.
\end{equation}

Collecting terms in $A$ and $B$ and dividing through by the coefficient of $B$, one finds that

\begin{equation}
    B = C_1 A.
    \label{Bcoeff}
\end{equation}

We can now express all other state populations in terms of $A$. As a final constraint, we impose the condition that the sum over all state population fractions must be equal to 1. Expressing all state populations in terms of $A$ and setting their sum equal to 1 yields equation (8) of the main article. Combining this with equations (\ref{F}) to (\ref{C}) and (\ref{Bcoeff}) above directly gives all of the steady state population fractions.

\section{Numerical estimation of second derivatives}

Before gradient estimation is performed, linear interpolation is used to obtain, from our irregularly-spaced raw data points, data corresponding to the fractional absorption of each laser beam at each point on a regular, 2500 $\times$ 2500 point grid, running between the minimum and maximum frequencies of each laser. Once this regularised data is obtained, each first derivative is estimated using a symmetric, linear estimation based on the values at grid points ten spaces in either direction of the point in question, such that, for example
\begin{equation}
     \left. \frac{dS_1}{dx} \right|_{x,y} \approx \frac{S_1(x+10,y)-S_1(x-10,y)}{20},
\end{equation}
where $x$ and $y$ correspond to the indices of the grid points, such that $dS_1/dF_{\mathrm{cooler}}$ etc. are ultimately found by multiplying the corresponding $dS_1/dx$ values by the frequency spacing between points on the corresponding axis. Equivalent expressions clearly exist for the three remaining first derivatives.

The calculated first derivatives are saved as arrays. This enables the corresponding second derivatives to be calculated, from the first derivative estimates, using exactly the method shown above.

This is a simplistic and un-optimised method of numerical derivative estimation, and more advanced techniques are certainly available. However, it was intentionally chosen in order to more accurately simulate the kind of low-complexity calculation that is likely to be performed in real time by experimental hardware, and to therefore better reflect a realistic implementation of our technique in an actual technological application. The use of more advanced gradient estimation techniques, potentially with a lower susceptibility to experimental noise, would likely improve upon the results given here but might overestimate the performance of the technique in realistic experimental implementations.


\section{Complications of a full, theoretical model}

In addition to requiring a quantum master-equation approach to achieve high accuracy under all conditions \cite{satspec_theory}, such a model would have to include a full analysis of transient, non-equilibrium effects in atoms that traverse the (spatially non-uniform) laser beam, integrated over all possible traversal speeds and trajectories in the thermal atomic vapour. It would also need to allow for the fact that the approximations typically made when analysing saturated absorption spectroscopy --- that the `probe' beam is weak and the `pump' beam is strong \cite{satspec_theory,preston1996doppler, pumpspec1} --- are not valid in this system. It would have to correctly handle the attenuation of the laser light as it passes through the atomic vapour, including the fact that the returning beam's intensity will depend on the attenuation of the outward beam, which in turn depends on the return beam's properties. The lifetime of the hyperfine states of the 6S$_{1/2}$ manifold against collisional redistribution can not be neglected, even for atoms that leave the beam and might return to it later, necessitating an in-depth analysis of both interatomic collisions and collisions with the walls of the cell. All of the above effects can have non-negligible consequences for our results, and the authors do not believe the omission of any to be a good approximation in realistic systems of this kind. In short, the complexity of a model capable of providing accurate quantitative agreement with our results is very high, and since such a model is not necessary to support our conclusions or elucidate the underlying principles we leave it for future work.


\section{Application within small vapour cells in miniaturised devices}

Figure 3(a) of the main article was taken using a 75\,mm long vapour cell. This size of cell is typical of existing, lab-based experiments but is not plausible in the next generation of miniaturised technologies for use outside the lab, or for spectroscopic devices that require a high spatial resolution. A smaller cell necessarily leads, for a given vapour pressure, to a lower 2D (projected) atom density and thus to a lower optical depth for laser beams passing through the cell. In addition, the experiment for which we give results was performed with Cs atoms, which have the highest vapour pressure of any commonly-used alkali metal species at 298\,K. 

As a result of the high total optical absorption in our experiment, our results --- in particular figure 2(a) --- do not make the degree of enhancement of the optical absorption via the use of dual-frequency spectroscopy immediately obvious. In order to see how large an enhancement of the spectroscopic signal is likely to be possible when using a lower total optical depth, it is necessary to view the results in logarithmic space, thus illustrating the strength of the response per unit of atomic density. Supplementary figure 1 (below) shows the same results as given in figure 2(a) of the main article, but plotted on a logarithmic scale. Here, the degree of benefit offered by this technique in portable devices with small vapour cells, or in any other situation with reduced total optical depth, is clearly evident. 

\begin{figure}[h!]
\centering
\includegraphics[width=0.7\linewidth]{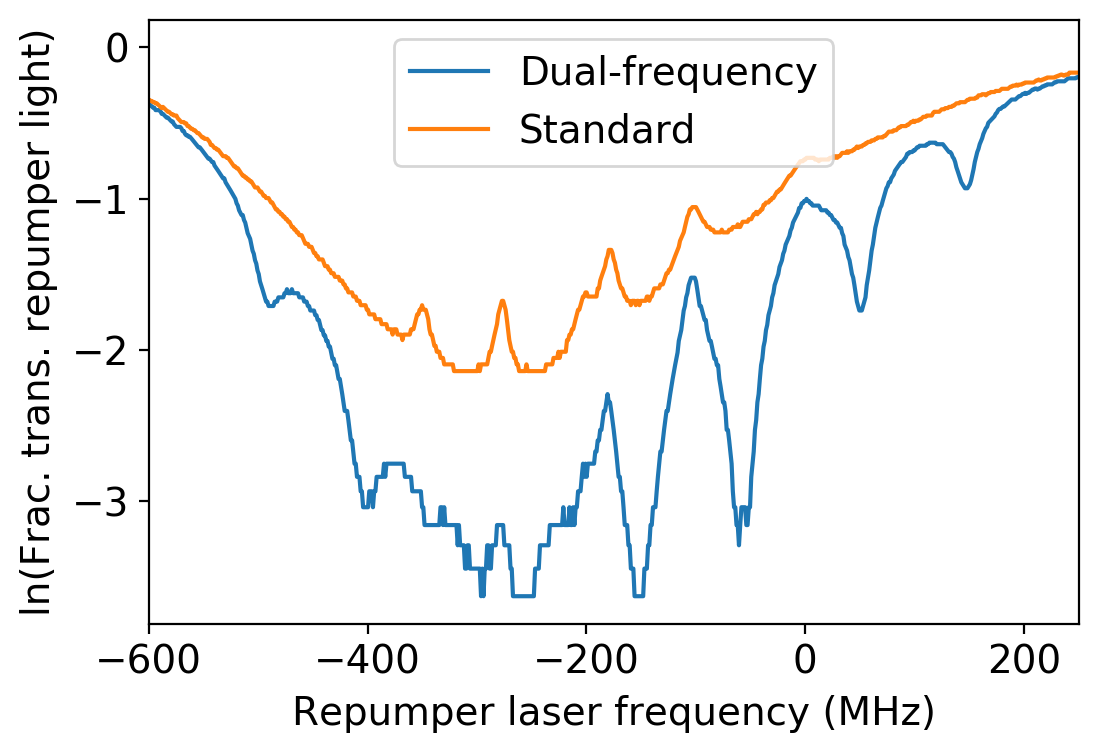}
\caption{Natural logarithm of the spectroscopy signal from ``Photodiode Repumper" (see figure 2(a) of main article) while light from the cooler laser, tuned $365$\,MHz below the $F=4 \rightarrow F'=5$ transition, is also present in the cell (blue line). For reference, a standard saturated absorption spectroscopy signal from the same apparatus (taken by blocking the light from the cooler laser) is also shown (orange line). Repumper laser frequency is given relative to the F=3$\rightarrow$F$'$=4 transition.}
\label{fig:cutthroughs}
\end{figure}

\section{Spectroscopy with injected light from an independently stabilised laser}

In the main article, we focussed on the more common experimental scenario of wanting to frequency-stabilise two lasers simultaneously. However, here we consider the possibility --- which may be of use in certain applications --- of using injected light from an independently stabilised laser to enhance the spectroscopic stabilisation signal for a second laser.

In this case the SLSF is determined by two factors: the sensitivity of the spectroscopic feedback signal to changes in the frequency of the target laser about its `lock point' (the frequency at which it is stabilised), and the scale of unwanted shifts in the lock point of the target laser that will result from changes in the frequency of the injected light. Since there is no reason for these two sources of instability to be correlated they can be added in quadrature when determining the likely magnitude of the laser's frequency deviations from its desired value.

If we assume that the injected laser is stabilised using standard saturated absorption spectroscopy we can label the likely scale of its frequency deviations from its setpoint as $\delta F_0$. Locking based on saturated absorption spectroscopy and current modulation generates a response signal $S$ approximately proportional to the rate of change of the saturated absorption signal voltage with laser frequency:
\begin{equation}
    S \approx C \frac{dV}{dF},
\end{equation}
where C is a system-specific proportionality constant, F is the laser frequency and V is the output voltage of the photodiode following the spectroscopy apparatus, normalised against its maximum value when the laser light is far off resonance and there is negligible absorption in the atomic vapour. The parameter of interest --- the sensitivity of the response signal to changes in laser frequency --- is therefore given by
\begin{equation}
    \frac{dS}{dF} \approx C \frac{d^2V}{dF^2}.
\end{equation}

\begin{figure}[h!]
\centering
\includegraphics[width=0.7\linewidth]{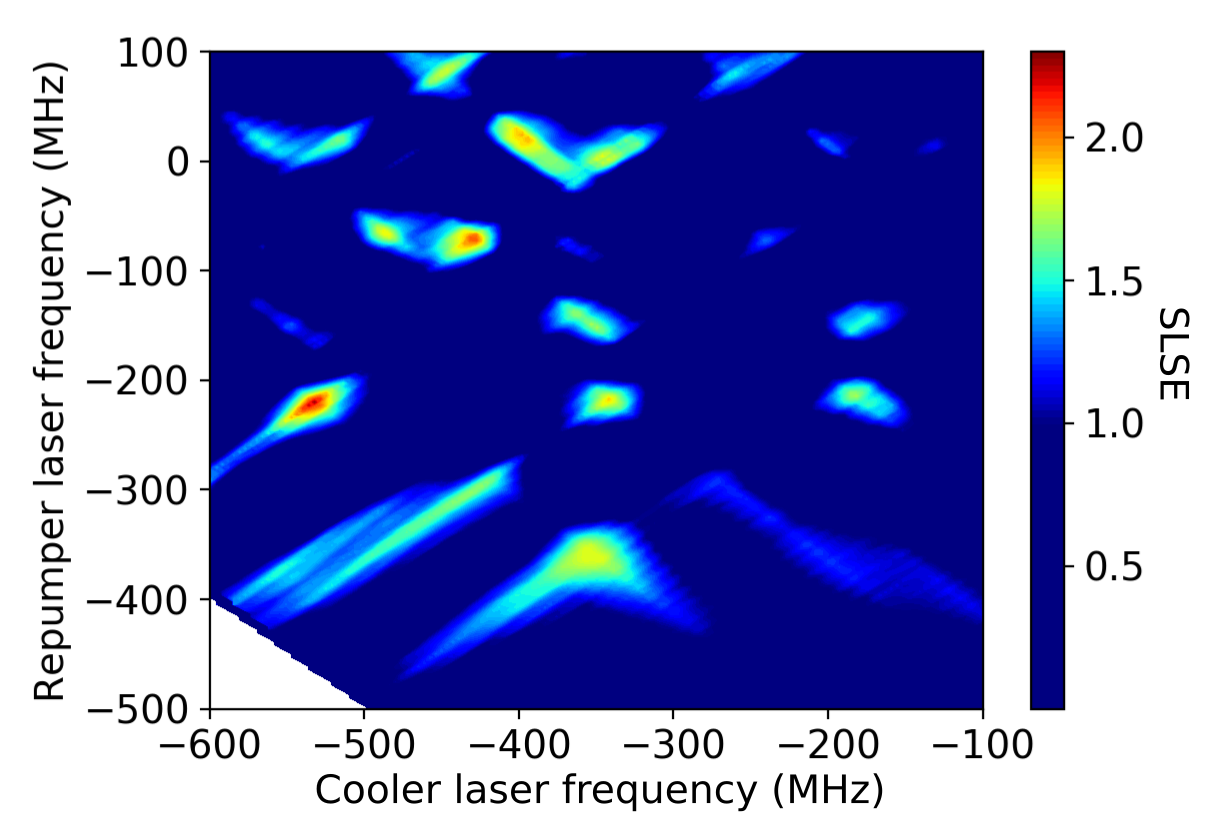}
\caption{Signal-limited stability enhancement (SLSF --- see main text for full definition) resulting from laser frequency stabilisation based on dual-frequency spectroscopy of the repumper laser, with injected light independently stabilised close to resonance with the cooler transition, compared to the maximum achievable using conventional saturated absorption spectroscopy. The SLSF is plotted as a function of both laser frequencies. Frequencies for the cooler and repumper lasers are given relative to the F=4 to F$'$=5 and F=3 to F$'$=4 transitions respectively. The results clearly show the potential advantages of this form of spectroscopy.}
\label{fig:dualstab2}
\end{figure}

To compare the efficacy of dual-laser spectroscopy to standard saturated absorption spectroscopy, we therefore compare the maximum values of $d^2V/dF^2$ measured with and without the second beam present in the vapour cell, defining the ratio between them as $R$. This ratio reaches values $>$3 for our experimental data. However, we must also take into account the effects of deviations in the injected light's frequency, giving
\begin{equation}
    \label{injres}
    \delta F \approx \sqrt{(\delta F_0 R)^2 + (\delta F_0 Q)^2},
\end{equation}
where $Q$ is defined by
\begin{equation}
    Q =  \frac{d^2V}{dF~dF_{\mathrm{injected}}} \left/ \frac{d^2V}{dF^2} \right. ,
    \label{Q}
\end{equation}
and we have assumed that the two different contributions to laser frequency variation --- deviation from the lock point and lock-point mobility arising from variations of the injected laser frequency --- are independent and can be added in quadrature. The derivatives of photodiode voltage with respect to $F$ in (\ref{Q}) are assumed to be evaluated at the chosen lock point and in the presence of the injected beam. 
We denote our figure of merit, the SLSF, by $E = \delta F_0/\delta F$. Figure \ref{fig:dualstab2} shows a colour plot of the experimentally derived values of the SLSF for the repumper laser with cooler light injection, as a function of both laser frequencies. Note that the experimentally derived SLSF value requires numerical estimation of local gradient functions, and that the exact method by which these are calculated can have some influence on the result --- see \S2 above for more details. It can be seen that there are numerous regions in the plane where $E>1$, with the maximum value of $E$ equal to 2.26. This demonstrates the clear advantages of the technique in terms of laser stability.

\section{Trapping and cooling Lithium atoms: an important application area}

The technique may prove particularly useful for spectroscopy of atomic Lithium. Lithium reacts chemically with most standard glasses, meaning that vapour cells for Lithium are typically much larger and more costly than those for other alkali metals; the benefit of using only one vapour cell to stabilise two lasers is correspondingly enhanced. Furthermore, the absence of a `cycling transition' in atomic Lithium, brought about by the small energy separation of the hyperfine states of the upper manifold of the D2 line, suggests that the benefits of avoiding optical pumping to dark states may be further enhanced for Lithium spectroscopy.

\section*{References}
\bibliographystyle{iopart-num}
\bibliography{sample}